\documentclass[twocolumn,showkeys,preprintnumbers,amsmath,amssymb,pre]{revtex4} 

\usepackage{graphicx}
\usepackage{dcolumn}
\usepackage{bm}
\usepackage{color}

\newcommand{\bq}      {\begin{eqnarray}}
\newcommand{\eq}      {\end{eqnarray}}
\newcommand{\rf}[1]    {(\ref{#1})}

\newcommand{\cR}       {{\cal R}}
\newcommand{\lp}       {\left}
\newcommand{\rp}       {\right}

\newcommand{\df}       {{\rm d}}

\newcommand{\nt}       {\noindent}

\begin{document}


\title{Highly under-expanded jets\\ in the presence of a transverse density gradient}

\author{Marco Belan}
\author{Sergio De Ponte}%
\affiliation{%
Politecnico di Milano, Dipartimento di Ingegneria Aeronautica e Spaziale\\
Campus Bovisa Sud - Via La Masa, 34 - 20156 Milano - Italy
}%

\author{Daniela Tordella}
\affiliation{
Politecnico di Torino, Dipartimento di Ingegneria Aerospaziale\\
corso Duca degli Abruzzi 24 - Torino - Italy
}%
%


\date{\today}

\begin{abstract}
An experimental research concerning highly underexpanded jets made of different gases from the surrounding ambient is here described.
By selecting different species of gases, it was possible to vary the jet-to-ambient density ratio in the  0.04 to 12 range and 
observe its effect on the jet morphology. 
By adjusting the stagnation and ambient pressures, it has been possible  to select the Mach number of the jets, independently from the density ratio.
Each jet is therefore  characterized by its maximum Mach number, ranging from 10 to 50.
The Reynolds number range of the nozzle is $10^3$ to $5\cdot 10^4$. 
The spatial evolution of the jets was observed over a  much larger scale than the nozzle diameter. 
The gas densities were evaluated from the light emission induced by an electron beam and the gas concentrations were obtained by analyzing the color of the emitted light. The results have shown that the  morphology of the jets depends to a greater extent on the density ratio. Jets that are lighter than the ambient exhibit a   more intense jet-ambient mixing than jets that are heavier than the ambient, while the effects of changing the jet Mach number do not seem to be too large in the explored range. These results can be expressed by means of two simple scaling laws relevant to the near field (pre-Mach-disk) and the mid-long term field (post-Mach-disk), respectively.
\end{abstract}



\maketitle

\section{Introduction}


The experiment here described has been  devoted to the study of  hypersonic underexpanded jets.  
In the last few decades, these jets have received a great deal of attention  because of their importance in several fields such as basic fluid dynamics and astrophysics, and in applications for the aeronautical and mechanical industries. 

The near field of underexpanded supersonic jets is characterized by a well known structure constituted by an intense expansion up to a high Mach number which ends in a normal shock, the Mach-disk, and which is surrounded by a barrel-shock. This structure was  described in the early theoretical and experimental works  by Ashkenas and Sherman \cite{ashk}, Crist et al. \cite{cris}. Later on, the high expansion present in these jets was represented by models that linked the Mach numbers and shock properties to the entropy change of the fluid from stagnation to the final after-disk state \cite{youn}. 

Very high pressure ratios can also give rise, at the end of the expansion, to a transition from a continuum to an almost collisionless flow. The relevant nonequilibrium phenomena have been the subject of numerous investigations, see the work of Cattolica et al. \cite{catt},
who made kinetic temperature measurements and qauntified the anisotropic partition of the random thermal energy along the direction parallel and normal to the mean flow.

The properties of underexpanded jets have  also been considered in astrophysics, where the evolution of jets observed in young stellar objects is an important issue  \cite{norm}. In this field, some information  may emerge from the study of the underexpanded jet morphology as a function of the Mach number and jet-to-ambient density ratio \cite{bel1}. 

Numerous investigations have focused on underexpanded jets, and there is an extensive literature on the subject which includes  compared numerical simulations and laboratory experiments (for example \cite{damb,wilk}), both in basic and applied research. 
In general, the near jet is described in a detailed way, and complete information is given on its velocity, pressure, temperature and density field 
\cite{nish,mcda}, but most  of the literature on these flows does not report data about the subsequent part of the jet, although there are a few exceptions 
\cite{norm,wilk}. Moreover, it is not easy to find data about the dependence of the jet properties on the jet-to-ambient density ratio. 

The experimental facility employed here is suitable for investigating the spatial evolution of a jet over a scale
of nearly two orders of magnitude larger than the formation scale, which is here defined as the nozzle throat diameter. 
This makes it possible to investigate a long region downstream from the Mach disk. Furthermore, a unique feature of the present experiment is the possibility of studying the effects of two flow control parameters, the stagnation-to-ambient pressure ratio and the jet-to-ambient density ratio, independently of each other. The measurement techniques are based on the electron beam technique. The fluorescence induced in plane sections of the flow field are acquired as digital images and processed by means of a  specially developed algorithm, which yields density and species concentration maps \cite{bel3}. 

A number of jets has been  studied here by varying  the stagnation-to-ambient pressure ratio 
(which is the compressibility parameter that sets the maximum Mach number and the Mach disk location in the near jet) and the jet-to-ambient density ratio over a  range. The variation of this last paramenter was possible  due to the  choice of  different pairs of gases from  a set of a few noble gases, such as Helium, Argon or Xenon (this is the parameter that sets the transverse density gradient in the intermediate and far jet). The density and concentration fields were determined for each jet, and the relevant Reynolds number was carefully characterized. 
It has been shown that the ambient density conditions play an important role on jet dynamics and morphology. The obtained
results could help us to understand their stability and transport properties. In particular, it has been shown that light or {\em underdense} jets (where the jet density is less than that of the ambient) spread out faster than  heavy or {\em overdense} jets. This is in agreement with previous  findings relevant to finite-thickness one stream compressible shear layers with uniform density (Gropengiesser, 1970) \cite{gropen} and to compressible turbulent shear layers (Papamoschou and Roshko, 1988)\cite{papa}. In the present study, we have been able to link the spreading angle before (near field) and after (mid-long term field) the Mach-disk and the jet/ambient density ratio by means of two simple scaling laws.

This work is organized as follows: section II presents the flows under study and the laboratory facilities, section III describes the data analysis methods, section IV contains the experimental results and section V the relevant discussion.

\section{Jets under study and laboratory set-up}

\begin{figure}[ht]
\centering
\includegraphics[width=\columnwidth]{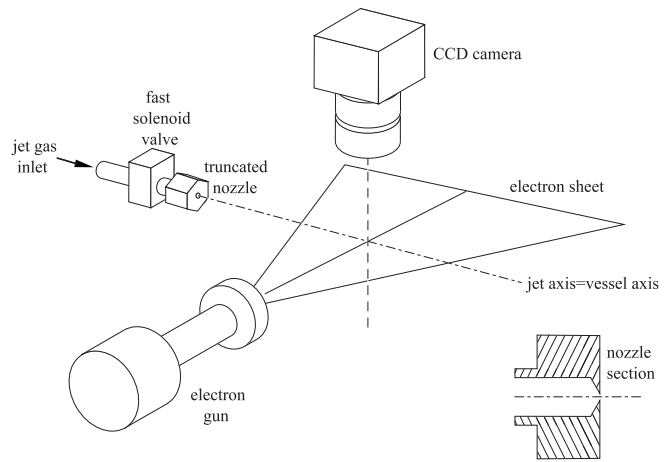}
\caption{\label{setup} Experimental setup. The ambient gas inlet is mounted onto the vacuum vessel wall.
The nozzle section is shown in the lower right corner.}
\end{figure}

The present experiments are carried out in a system that was specifically designed for the study of hypersonic jets \cite{bel1,bel3}. The system facilities consist of a cylindrical vacuum vessel equipped with suitable nozzles, and an electron gun that makes it possible to obtain the fluorescent plane sections of the jets, which have here acquired as digital images using  a  CCD color camera. The setup inside the vessel is sketched in fig. \ref{setup}. 

Here the tested flows are strongly underexpanded jets obtained from a truncated sonic nozzle that flow along 
the longitudinal axis of the vacuum vessel. The vessel has a diameter of 0.5 m which is always 5 times larger than the diameter of the jets, which ranges from 0.01 to 0.1 m. 
The vessel is equipped with 2 valve systems that control the jets issued by the nozzle and the ambient gas;
this offers the opportunity of using different gases for the jets and the surrounding ambient. 
The jet stagnation pressure $p_0$ and the ambient pressure $p_{amb}$ are determined by adjusting the opening times of the 2 valve systems, 
using by a suitable electronic device. The essential data of the experimental system are shown in table \ref{tab}.

\begin{table}
\begin{center}
\caption{\label{tab} Capabilities of the experimental system.}
\renewcommand{\arraystretch}{1.5}
\begin{ruledtabular}
\begin{tabular}{lcc}
Vessel length 				& 	& 2m to 5m (modular) \\
Vessel diameter 			& 	& 0.5m \\
Vessel (ambient) pressure 		& $p_{amb}$ & 1.5 to 200 Pa @ 300 K \\
Nozzle throat diameter 			& $D$ 	& 2mm \\
Jet stagnation pressure 		& $p_0$ & 2000 to $2 \cdot 10^5$Pa \\
\begin{minipage}[b]{2.8cm}
{Stagnation/ambient\\pressure ratio}
\end{minipage}
					& $p_0/p_{amb}$ & 10 to $1.3 \cdot 10^5$\\
Electron gun voltage 			& & up to 20kV  (16kV typ.) \\
Electron gun current 			& & up to 2mA   (1.5mA typ.)\\
\end{tabular}
\end{ruledtabular}
\end{center}
\end{table}

The accuracy of pressure transducers is 1\% on the $p_0$ measurements and 0.25\% on the $p_{amb}$ measurements.
The upper  ambient pressure limit is essentially set by according to the proper conditions for electron gun working, i.e. pressure below $200 \sim 300$Pa and 
density number no greater than $10^{23}$m$^{-3}$ in order to avoid excessive absorption of the electrons by the gases and light intensity saturation.  
The camera has a high sensitivity, 1Mpixel CCD, with a Bayer RGB filter, the operating mode is 12 frames/s for a 83 ms exposure 
or 24 frames/s for a 41 ms exposure. 

The flow time scale, defined as the ratio between jet diameter and local speed of sound, 
ranges from $10\mu$s in the nozzle throat to $1$ms at the end of the jet expansion, upstream from the Mach disk.
The typical jet outflow time is 0.5s, that is, hundreds or thousands of time scales, and  changes in the ambient can be observed in this lapse of time due 
to the jet/ambient mixing. However, due to the large mass flow of the primary pumps, fast variations in the ambient during the outflow time are avoided, and since the image exposure is sufficiently shorter than the outflow time, the large-scale properties of the jets show only very small variations over a single exposure, therefore the jets can be considered quasi-steady in each image. 

Choosing the jet/ambient pairs from Helium, Argon and Xenon and varying $p_0$ and $p_a$ in the allowed ranges
permits variations of the jet-to-ambient density ratio $\eta=\rho_{jet} / \rho_{amb}$, ranging from 0.04 to 12 in the near jet zone
and from 0.2 to 4 in the intermediate and far jet zone, to be obtained.

In these experiments, the Mach number of the jets in the near zone is very high, since the jet gas may approach
the limit velocity just upstream from the Mach disk, therefore $M$ values of up to 50 are possible.
The Knudsen number $K$
is always in the continuum regime, except for the particular case of helium jets with very high $p_0/p_{amb}$ ratios, where $K$ may approach 0.5 in a small region upstream from the Mach disk. It should be noticed that in this case and in general at the end of the expansion, the accuracy of the measurements
can deteriorate \cite{bel3}. In the following sections, the reported accuracies  always take  this problem into account.

\section{Data analysis}

\subsection{Density and concentration measurements}

Here the measurement method described by Belan et al. \cite{bel3} is used to obtain density and concentration maps 
of jet and ambient gases from the original digital color images.
The method is based on the analysis of the fluorescent emission $I$ from a gas excited by an electron beam,
which at the low pressures considered here 
can be expressed in the form
\bq
I=k\, n ,
\label{intens}
\eq
where $n$ is the gas number density and $k$ a proportionality coefficient.

The limits of validity of this approximated law are stated by the following conditions:\\
- low density ($n<10^{23}$m$^{-3}$), which ensures linearity and, for gas mixtures, a decoupled emission
(i.e. the total emission is simply the sum of the partial intensities due to different species)\\
- bounded range for the gas temperature ($70K<T<900$K), which cannot be so high as to have a visible emission,
or so low as to have an emission affected by nonequilibrium phenomena. 

These conditions have been widely satisfied by the tested jets, and in particular in the jet/ambient mixing zone.
The only exceptions are:\\
- a small region upstream from the Mach disk, where the temperature may fall below 70K and the uncertainties on the measurements can grow by more than  15\%.\\
- a small region close to the nozzle, where the density is too high and the emission is saturated.

The operations performed by the image post-processing algorithm are summarized hereafter, while the relevant details  
are reported in appendix \ref{ALG}. The  jet gas concentration along an image, $z_{jet}$,  is obtained reading the RGB values of each pixel that 
form the color ratio 
\bq
r = \frac{C_1}{C_2} = \frac{a_1 R + b_1 G + c_1 B}{a_2 R + b_2 G + c_2 B}
\eq
where the six coefficients $a_i,b_i,c_i$ are chosen on the basis of the gas pair involved in the experiment,
and using the relation
\bq \label{rel-Zg1}
  z_{jet} = \frac{k_{amb_1} - k_{amb_2} r}{k_{amb_1} - k_{jet_1} + \left(k_{jet_2} - k_{amb_2}\right) r}\;, 
\eq
where the four coefficients $k_{amb1},k_{amb2},k_{jet1},k_{jet2}$ are known from calibrations performed on pure gases.
The concentrations of the two gases are obviously linked by $z_{amb}=1-z_{jet}$. 
Once the concentrations are known, the density is given by the equation
\bq
\rho = C_1 \frac{z_{amb} m_{amb} + z_{jet} m_{jet}} {k_{amb1} z_{amb} + k_{jet1} z_{jet}}
\label{DENS}
\eq
where $m_{amb}, m_{jet}$ are the molecular masses of the gases in the ambient and in the jet, respectively.

\subsection{\label{METH} Measurements on plumes and mixing layers}

The aim of this section is to define suitable criteria to compare the different jets under study and to exploit their dependence on the control parameters. To this aim, the measurement method explained above will be used to determine the jet properties by means of density and concentration values.


\begin{figure*}[t!]
\centering
\includegraphics[width=\textwidth]{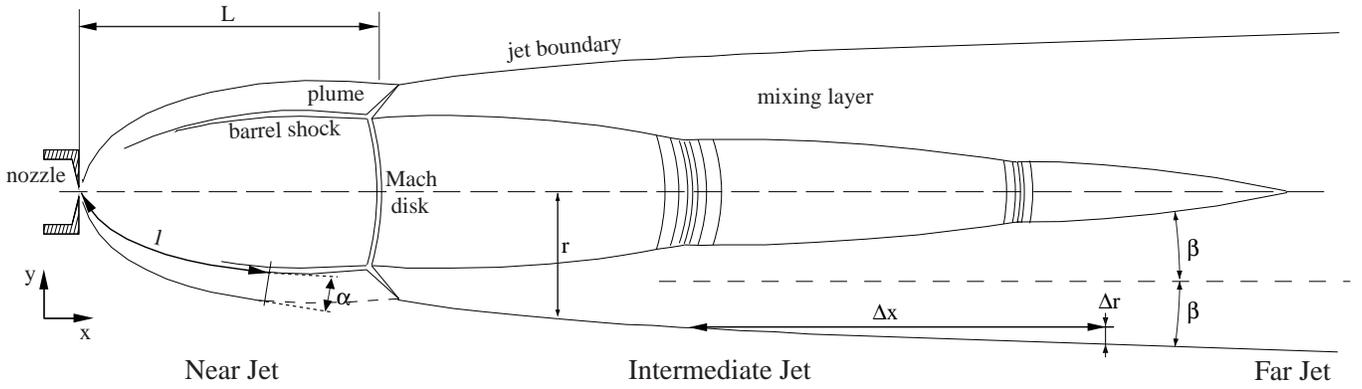}
\caption{\label{underexp}Structure of a strongly underexpanded jet. $L$ is the nozzle-Mach disk distance or barrel length, 
$\alpha$ is the plume growth angle measured on an arc of length $l$, $r$ is the conventional jet radius, and $\beta$ is the mixing layer spreading angle in the intermediate jet.}
\end{figure*}

At this point, it is important to consider the underexpanded jet structure in detail, referring to fig. \ref{underexp}.
In the short range upstream from the Mach disk, which is located at a distance $L$ from the nozzle (near jet or barrel zone), the Mach number increases to its maximum value $M_{max}$ just upstream from the normal shock, whereas the density ratio decreases very rapidly: this happens both for light and heavy jets. In the present case, the jet properties essentially depend  on the stagnation-to-ambient pressure ratio. 

\begin{figure*}[t!]
\centering
\includegraphics[width=\textwidth]{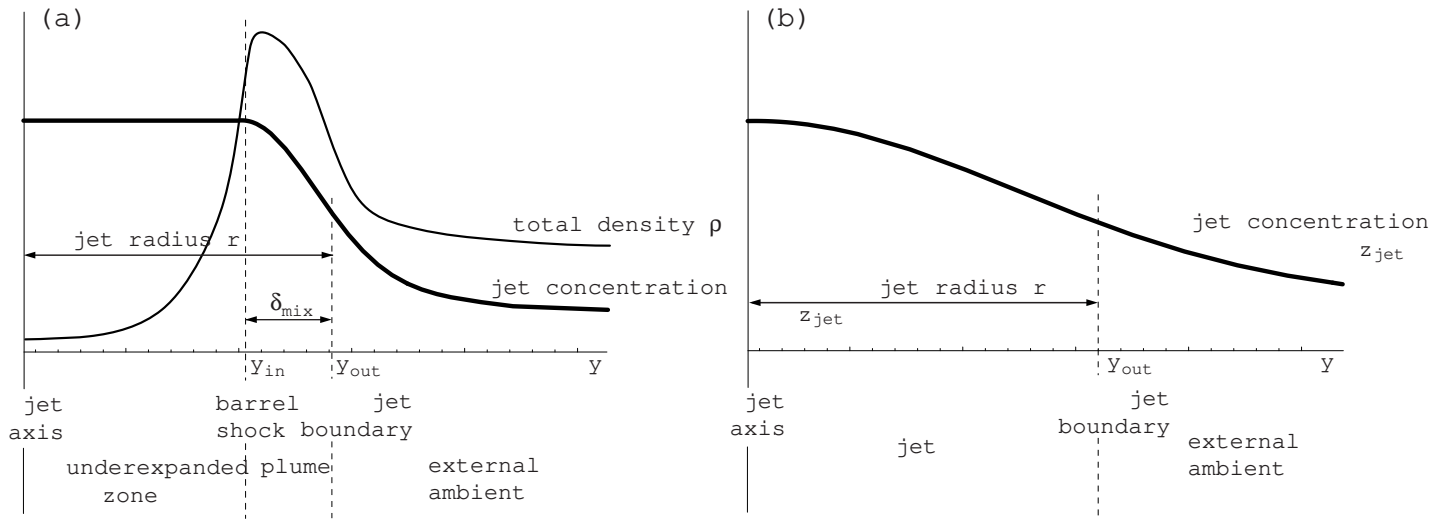}
\caption{\label{sect}Jet cross-sections in the near (a) and intermediate (b) regions.}
\end{figure*}

The typical trends of radial the concentration and density curves in this zone are shown in figure \ref{sect}a.
It should be noticed that $z_{jet}$ does not vanish in the outer zone: in fact, as the testing time increases in these experiments, the mass flow of the jet gas, which is not completely balanced by the primary pumps, despite their power, is partially mixed with the environment. In a given experiment, for example Ar in He, the Ar jet is therefore found  to travel in an environment of Ar mixed with He. Thus, in these experiments, the ambient contains an ineliminable part of the jet gas, but this is not a limit to the present research since this effect is directly taken into account in the density and concentration measurements.

Here, in the near jet, a 'mixing layer width' or 'plume width' can be defined as the radial distance between the barrel shock and the outer jet boundary
(figure \ref{sect}): the radial position of the barrel shock, $y_{in}$, can be identified as the point where the density curve has the maximum slope,
\bq
\lp( \frac{\df^2 \rho}{\df y^2} \rp)_{y_{in}} = 0
\eq
and the outer jet boundary can be identified by condition
\bq
\lp( \frac{\df^2 z_{jet}}{\df y^2} \rp)_{y_{out}} = 0
\label{jb}
\eq
i.e. where the derivative $\df z_{jet}/\df y$, with respect to the radial coordinate $y$, reaches its maximum value. It is easy to see that this is analogous to the classic 'half width at half height' criterion that is extensively  used for  jets in literature \cite{schl}.
The $y$ value where condition \rf{jb} is satisfied can also be taken as a conventional jet radius $r=y_{out}$, bearing in mind that the jet width $2\,y_{out}$, obtained as the distance between opposite boundaries, has a relative and not absolute meaning (for example, other quantities proportional to $y_{out}$ could be taken as conventional jet radii without any loss of generality when  similar jets are compared).

Therefore, in a given radial section, the plume width $\delta_{mix}$ can be defined as
\bq
\delta_{mix} = y_{out}-y_{in} .
\eq
For the sake of comparison, the jet cross section that should be considered when measuring $\delta_{mix}$ should be conventionally taken at the same axial position 
$X$ in each experiment, here,   $X=0.8L$, referring to the nozzle-Mach disk distance $L$ (barrel length), has been chosen; 
$X$ is not too close to the Mach disk zone to make the measurements too difficult. Once $\delta_{mix}$ has been determined, it is possible to calculate the plume growth angle $\alpha$: 
\bq
\alpha = \arctan (\delta_{mix}/l)
\label{alfap}
\eq
where $l$ is the relevant arc length, that is the length of the barrel shock between the nozzle and the measurement cross section for $\delta_{mix}$.  

In the long range (intermediate and far jet), downstream from the Mach disk, the Mach number remains quite low, and the jet properties  essentially depend on the density ratio. There are also density fluctuations in some underdense jets, which are shown in figure \ref{underexp} as thin line beams. Here, the mixing layer grows both inwards and outwards: it is possible to use the same definition as above for the outer boundary of jet and mixing layer, while  a criterion based on concentration and density values, even though technically possible, results to be too noise-sensitive for the inner boundary. Nevertheless, the long scale behaviour of the jets can be described by a spreading angle $\beta$, defined in terms of the outer boundary position.
This angle represents the increase in the mixing region in the intermediate field where an annular mixing layer surrounds the jet core;
in the final far field, where the jet is fully developed and the flow becomes uncompressible, $\beta$ instead represents  the spreading angle of the whole jet, but this field is beyond the scope of this work and will not be considered.

In order to identify the jet boundary in the intermediate field only the cross-sectional $z_{jet}$ curve is required; a typical trend is shown in figure 
\ref{sect}b. After choosing a pair of axial positions $x_1,x_2$ in the range $x>L$, the outer boundary is found using criterion \rf{jb}, and two values of the jet radius $r_1=y_{out}(x_1)$ and $r_2=y_{out}(x_2)$ are obtained. Then, the  spreading angle is obtained from the increments $\Delta r = r_2-r_1$ and 
$\Delta x = x_2 - x_1$ through relation 
\bq
\beta = \arctan (\Delta r/\Delta x)
\label{alfaj}
\eq
These measurements are usually performed choosing several $x_1,x_2$ pairs and calculating $\beta$ as a mean value. The $x_1,x_2$ pairs are choosen
in the  $x>1.2L$ range to avoid setup effects on the mixing layer.

In order to ensure the proper evaluation of the concentration and density curves, together with their derivatives, all the images analyzed in this work have been filtered to improve the signal-to-noise ratio by means of standard local averages in the small neighborhoods of each pixel. The non-uniformity of the electron sheet in
the images is also compensated for. 

\subsection{\label{RE} Reynolds number}

The description of the tested jets  should be completed with their Reynolds numbers.
Here,  the longitudinal Reynolds number  is used for this purpose, and is defined in the barrel zone ($x<L$) as:
\bq
Re_x = \frac{ U x \rho }{\mu}
\label{rex}
\eq
where $U = M c = M(x) \sqrt{\gamma \cR T(x)}$ is the centerline velocity,
the local Mach number $M(x)$ is calculated according to Ashkenas et al. and Young \cite{ashk,youn} and $T(x)=T[M(x),T_0]$ is the isoentropic temperature relevant to the stagnation temperature $T_0$ and to $M(x)$.
The density $\rho$ is measured directly on the images, and $\mu(T)$ is calculated for each gas by means of the Sutherland law where possible and of suitable models at the lowest temperatures \cite{kest,holl,gris,macr}.
For the sake of comparison among different jets, in what follows for each jet it will be reported the Reynolds number attained just upstream of the normal shock ($x\sim L$). 

A remarkable property of these flows is the range of Reynolds numbers covered by each jet; actually in these jets the Reynolds number, whether diameter based or jet-length based, falls off suddenly after the Mach disk. This variation can be of several orders of magnitude, since the strong temperature rise increases the viscosity dizzy, while the velocity decrease is smaller and the space scale varies slowly. The same problem arises if the Reynolds number is estimated for the mixing layer zone. An example can be given for a typical Argon jet having a pressure ratio $p_0/p_{amb}=10^4$: in this case, the diameter based Reynolds number is $2.4 \cdot 10^4$ in the throat, then it increases up to $6.7 \cdot 10^5$ upstream of the Mach disk, and falls off to about 83 downstream.

\section{Results}

The present results have been obtained considering 280 image sequences, each  consisting of 18 or 36 images of jets under quasi-steady conditions; a wide range of Mach numbers and density ratios, ranging from underdense jets to overdense jets,  is therefore available. Among these images, those with the best S/N ratios were selected to obtain the presented results. In what follows, the image processing techniques are first applied to a selected sample image as an example; the general results are then presented.

\subsection{An image processing example}

\begin{figure}[h]
\centering
\includegraphics[width=\columnwidth]{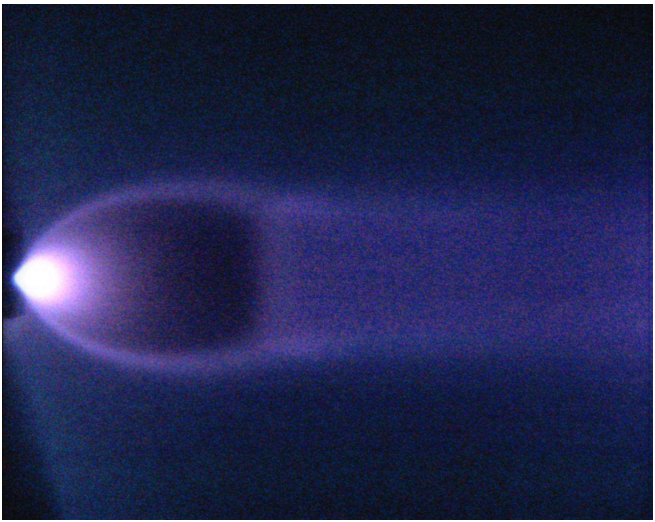}
\caption{\label{XeHejet}Example: Xenon jet in a Helium medium. Pressure ratio $p_0/p_{amb} \sim 4.3 \cdot 10^3$,  
Mach before the normal shock $\sim 41$, Reynolds number at the nozzle exit = $2.5 \cdot 10^4$.}
\end{figure}

Figure \ref{XeHejet} shows an Xenon jet flowing in a Helium ambient. The jet stagnation pressure is 
$p_0=(4.10 \pm 0.20)\cdot 10^4$ Pa, while the pressure of the Helium ambient is $p_{amb}=(9.86 \pm 0.1)$ Pa.
The figure is overcontrasted for the sake of clarity. The measurements have been performed along the axis and on different cross-sections (see below).

\begin{figure}[ht!]
\centering
\includegraphics[width=\columnwidth]{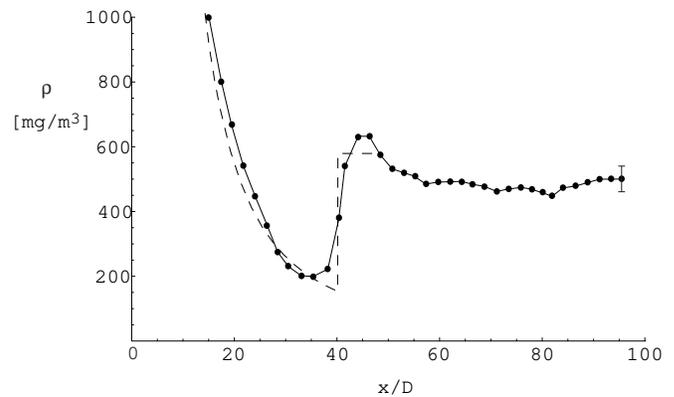}
\caption{\label{XeHerx}Axial density for the Xenon jet visualized in fig. \ref{XeHejet}. A sample error bar is shown on the right.
The zone on the left  cannot be analyzed because of the high density values (saturated output). 
The dashed line shows, as a reference, an equivalent isentropic expansion followed by a normal shock.} 
\end{figure}

\begin{figure}[ht!]
\centering
\includegraphics[width=\columnwidth]{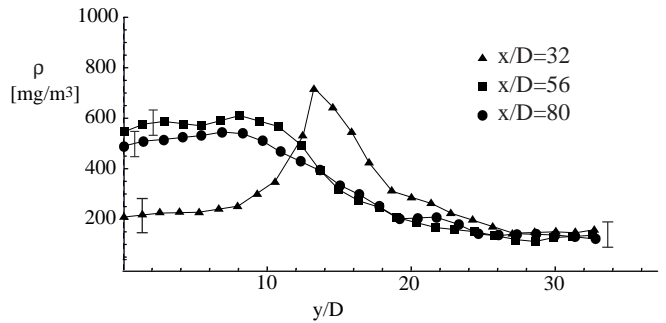}
\caption{\label{XeHery}Cross-sectional density curves for the Xenon jet visualized in fig. \ref{XeHejet}. Sample error bars are shown on both sides of each curve.}
\end{figure}

Figures \ref{XeHerx} and \ref{XeHery} show examples of the axial and cross-sectional density and concentration measurements performed in fig. \ref{XeHejet}. Each figure contains sample error bars, as typical values of the uncertainties, which are obtained by propagation of the errors, accounting for noise values along the image and for  calibration coefficient uncertainties in the image processing algorithms. It can be seen, in fig. \ref{XeHerx}. that the density values along the jet axis agree well with the relevant theoretical isoentropic expansion, a fact which should presumably be related to the heavy weight of the Xenon atoms. However, this is not a general property of strongly underexpanded jets, since the gap between the measured and isoentropically calculated values for underdense jets, for example He in Ar, is larger \cite{bel3}. The Mach disk position is in good agreement with Young's  theoretical estimation \cite{youn}.

Three radial density curves can be seen in fig. \ref{XeHery} at different axial stations, namely $x=32D=0.8L$, $x=56D=1.4L$ and $x=80D=2L$,  
in terms of the initial diameter and the barrel length. Here the uncertainties are in general of the order of $\pm 10\%$ of the local axial density value, except in the first $x/D=32$ curve, where the axial zone has a lower S/N ratio, which is usual in a dark underexpanded zone. The barrel shock position is clearly visible in the  curve for $x/D=32$, the other two curves show the effects of species mixing in the considered range.  

\begin{figure}[ht!]
\centering
\includegraphics[width=\columnwidth]{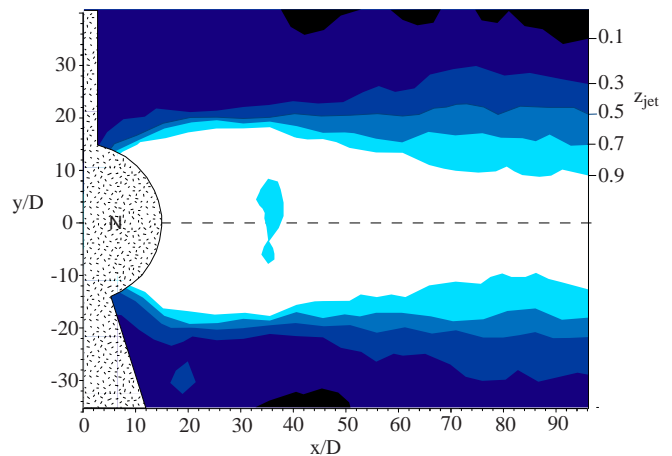}
\caption{\label{XeHezm}Concentration map of the Xenon jet visualized in fig \ref{XeHejet}.
The zone marked with "N" cannot be analyzed, since the relevant pixels are in the density saturation range, fall 
outside the electron sheet, or are influenced by the nozzle image.}
\end{figure}

Figure \ref{XeHezm} shows the Xe concentration map relevant to figure \ref{XeHejet}. Here, the almost  pure Xe zone ($z_{jet}>0.9$) slowly becomes
narrower as the jet travels along the axial direction. There is a visible spreading in the other contour lines, but the width of the region between the maximum variations in the $z_{jet}$ curve does not change fast, which means the jet boundaries spread out quite slowly according to rule \rf{jb}. However, in the following results it will be seen that the spreading of underdense jets is instead remarkably larger.

\subsection{Plume spreading in the near jet}

This section deals with results referring to the region between the nozzle and the Mach disk.
Among the numerous available jets, the ones selected according to the best S/N ratio are:\\

\nt
\small
8 Helium jets traveling in the Xenon ambient (labeled He/Xe)\\
13 Helium jets traveling in the Argon ambient (labeled He/Ar)\\
14 Argon jets traveling in the Helium ambient (labeled Ar/He)\\
8 Xenon jets traveling in the Helium ambient (labeled Xe/He)\\
\normalsize

Since the jet stream is accelerated in this zone,  each jet will be characterized by the Mach number 
$M_{max}$ at the end of the expansion, upstream from the Mach disk, for  comparison purposes among the jets. $M_{max}$ is calculated according to Young \cite{youn}. 
In an analogous way,  the maximum longitudinal Reynolds number $Re_x$, calculated as in \S\ref{RE} by equation \rf{rex} at the end of the expansion, will be reported for each jet.

\begin{figure}[h!]
\centering
\includegraphics[width=\columnwidth]{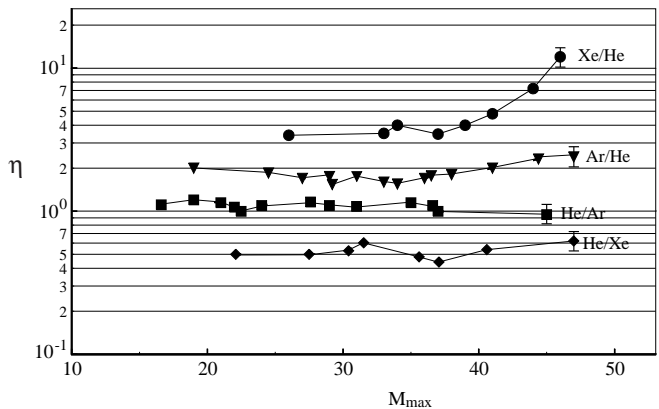}
\caption{\label{ratioM}Parameter domain $M_{max}\,\eta$ (maximum Mach number -- density ratio) of the tested jets. Sample error bars are shown on the right.}
\end{figure}

Since  the independent parameters in these experiments are the density ratio and the Mach number, each selected jet corresponds to a point
in the parameter domain $M_{max}\,\eta$, as shown in figure \ref{ratioM}. Some sample values of the uncertainties are shown on the right side of each curve,
since the parameter pairs $M_{max}\,\eta$ are not set directly because of to the experimental setup; their values depend on the pressure ratios 
$p_0/p_{amb}$ and the gas pairs, which are instead  set directly and which permit  the uncertainties of $M_{max}$ and $\eta$ to be calculated.
The reported density ratios $\eta$ refer to a cross-section lying at $0.8 L$ from the nozzle, that is, a value which is close to but not intercepting the Mach disk. The density ratio, on each cross-section, is obtained by measuring the maximum $\rho_{jet}$, i.e. the jet density at the outer surface of the barrel shock (the $\rho$-curve peak  in fig. \ref{sect}a) and measuring $\rho_{amb}$ in the outer far field. It is also possible to determine $\eta$ taking $\rho_{jet}$ on the jet axis, but this is less diagnostic, because the strong expansion in this zone makes the density ratio less dependent on the gas species, and makes the S/N ratio worse. For example, the overdense jet in figure \ref{XeHery} gives $\rho_{jet,max}/\rho_{amb}=4.83$, while 
$\rho_{jet,axis}/\rho_{amb}=1.35$ at the same cross-section.

\begin{figure}[h!]
\centering
\includegraphics[width=\columnwidth]{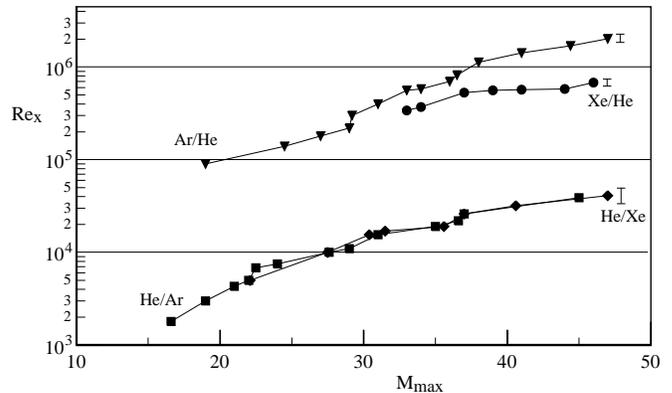}
\caption{\label{ReM} Maximum longitudinal Reynolds number $Re_x$ upstream from the Mach disk as a function of $M_{max}$. Sample error bars are shown on the right}
\end{figure}

All the jets considered here have nozzle Reynolds numbers, based on the throat diameter, ranging from 
$10^3$ to $5\cdot 10^4$.  For the selected jets, the maximum Reynolds numbers  attained upstream from the Mach disk are reported in fig. \ref{ReM}, as functions of the maximum Mach numbers. 

\begin{figure}[h!]
\centering
\includegraphics[width=\columnwidth]{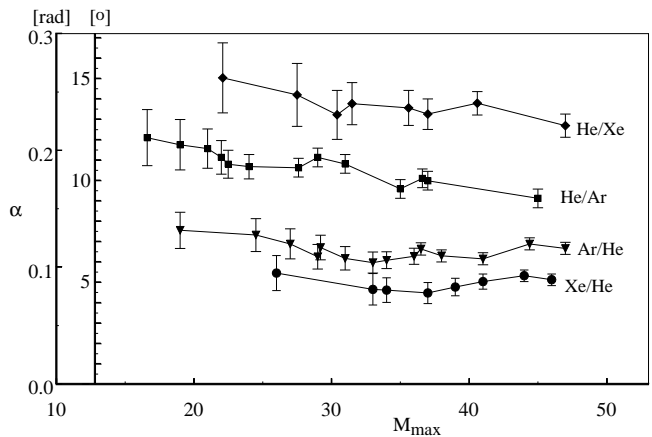}
\caption{\label{alfam}Spreading angles of the plumes outside the barrel shock vs $M_{max}$ for several jet/ambient gas pairs
The density ratio ranges are $0.50<\eta<0.62$ for He/Xe, $0.95<\eta<1.20$ for He/Ar, $1.54<\eta<2.40$ for Ar/He and $3.40<\eta<12.0$ for Xe/He.}
\end{figure}

Figure \ref{alfam} reports the growth angles of the jet plumes as functions of $M_{max}$ for different gas pairs, ranging over several density ratios. 
The relevant $\eta$ values, reported in the caption, are the same as those in figure \ref{ratioM}. The plume growth angles $\alpha$ are obtained from formula \rf{alfap}, as explained in \S\ref{METH}. The ucertainties are calculated by Gaussian propagation as above.

The general trend exhibited by these results is that underdense jets spread out more than 
overdense jets do. The plume spreading angle $\alpha$ seems to depend weakly on the Mach number $M_{max}$:  
it decreases slowly for underdense jets as $M_{max}$ grows, whilst  it becomes nearly independent for overdense jets.
The Reynolds dependence of $\alpha$ is also analogous, i.e.  $\alpha$ diminishes slowly for underdense jets as $Re_x$ grows.
Apart from these properties, it is worth noting that  the density gradients in overdense jets are sharper , and it can be shown that  the shocks are also thinner, a result that has been confirmed by acquiring monochromatic intensified images with very short exposures \cite{bel2}.

\begin{figure}[h!]
\centering
\includegraphics[width=\columnwidth]{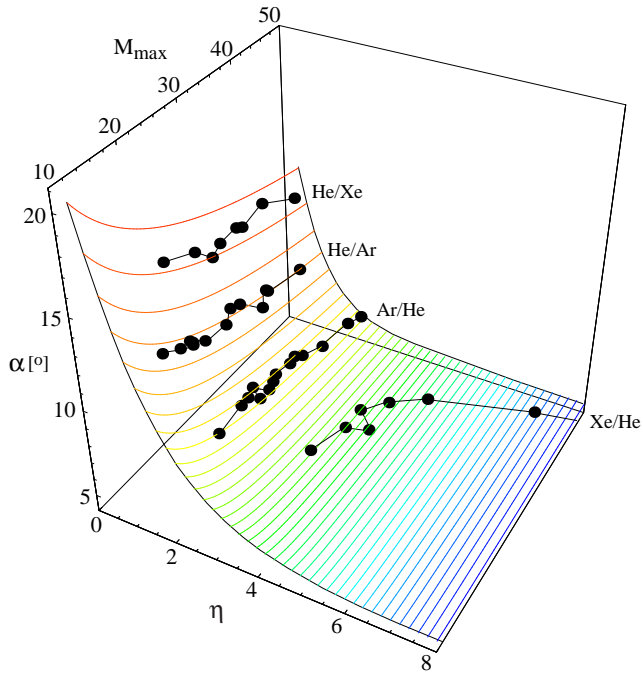}
\caption{\label{a-scale}Scaling law \rf{sca} for the plume spreading angle as a function of the Mach number $M_{max}$ and density ratio $\eta$.}
\end{figure}

The global trend of the $\alpha$ values can be expressed by means of a scaling law in the form
\bq
\alpha= a + b\,M_{max}^{(n + k \eta)};
\label{sca}
\eq
this law best fits the experimental data  with the coefficients $a=4.67,\, b=36.9,\, n=-0.235,\, k=-0.315$, which give an estimated error variance of 0.627. The relevant surface $\alpha(M_{max},\eta)$, superimposed onto the experimental data, is shown in figure \ref{a-scale}.

\subsection{Mixing layer spreading in the intermediate jet}

\begin{figure}[h!]
\centering
\includegraphics[width=\columnwidth]{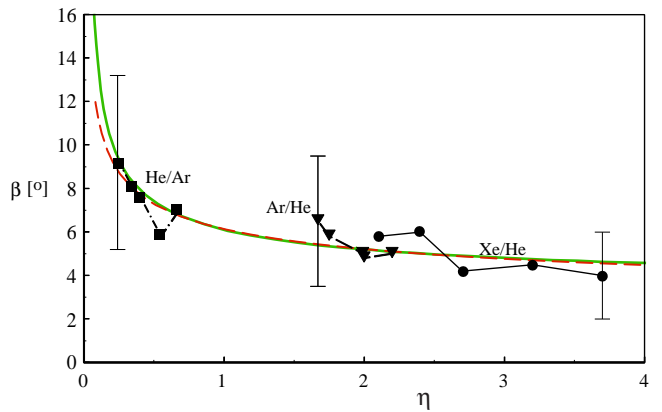}
\caption{\label{betafar}Spreading angles of intermediate jets vs density ratio (the axial Mach number downstream from the Mach disk is about 0.45 for all the jets). The green line is a fit according to  Papamoschou and Roshko (1988) (the red dashed line is the best scaling law fit, see formula \rf{scf}.} 
\end{figure}

This section contains results referring to the intermediate jet region, downstream from the Mach disk. 
The mixing layer spreading angles $\beta$ for the intermediate jets are obtained from formula \rf{alfaj}, as explained in \S\ref{METH}. These angles are determined from data collected on an axial domain $1.2 L<x<x_{max}$, where $x_{max}$ ranges from $2.5L$ to $5L$, depending on the jet under test, and avoiding positions where the mixing region reaches the jet axis.
Figure \ref{betafar} reports the $\beta$ values as a function of the density ratio, where the very weak dependence on the Mach number has been omitted (see below). The $\eta$ values are determined from the axial mean value of the jet centerline density, which varies slowly in the considered range, and from the density value of the surrounding far ambient. 
These measurements can be very sensitive to  image noise, depending on the involved gases, and require a wide image field: for these reasons, figure \ref{betafar} just reports  a subset of the available jets.
The uncertainties have been calculated with the same method as above; the resulting values are not small.

Here the dependency on the Mach number is weak: the Mach number $M_1$ on the jet axis after the normal shock is in fact always subsonic, and very close to the minimum theoretical value ($.448<M_1<.455$), since the flow upstram of the normal shock has $M_{max}>10$ for all the jets studied here. After the normal shock,  the jet pressure is also very close to the ambient pressure. Thus, the compressibility effects in this region seem to be small and the jet evolution is essentially determined by the density ratio. 

Nevertheless, although the jet core becomes subsonic and nearly incompressible after the Mach disk, 
an annular supersonic region in the outer jet may survive over  a short distance, which is less than $2L$ in these experiments.
This is due to the different behaviour of the inner and outer streamlines: the inner streamlines fall from $M_{max}$ to $M_1 \sim 0.45$ after crossing the Mach disk, whereas the outer streamlines that cross the barrel and the reflected shock visible in fig. \ref{underexp}
-- both obliques -- have a larger exit Mach number 
than unity (see, for instance, Norman and Winkler\cite{norm}). Here the flow remains supersonic until the lateral mixing region expands over the entire jet section,
and this can give rise  to further expansions along the jet up to the sonic condition,  followed by recompressions \cite{klav}. These phenomena were in fact observed in the underdense He/Ar and He/Xe jets at pressure ratios of $100<p_0/p_{amb}<1000$.

On the other hand, if the effects of the supersonic tongues and the possible velocity oscillations on the axis,
in the intermediate jet zone are disregarded, the flow can be considered as a core stream surrounded by an annular mixing layer, with the static pressures of the core and ambient very close to each other.  Under these hypotheses, and expressing  the mixing layer thickness measured along the $x$ axis with respect to a given origin $x_0$, as $\delta(x)$, the growth rate 
\[
\delta'(x)=\frac{\delta(x)}{x-x_0}
\]
can be estimated  by the measured angle $\beta$, since $2\, \beta \sim \arctan \delta' \sim \delta'$, under the hypothesis of small $\beta$.
The measured angles can be compared with the Papamoschou and Roshko model \cite{papa}, which can here be expressed by a law of the kind
\bq
\beta = c\, [ 1 + \eta^{-1/2}] ,
\label{prfit}
\eq
where $c$ is a constant. The best fit based on this law has a coefficient $c=3.01$, see formulas 14 and 15 in \cite{papa}. The resulting curve is shown in figure \ref{betafar} together with the present experimental data.

The general trend in the intermediate and far regions may now be outlined as follows. In the underdense jets, the mixing layer spreads fast, even toward the jet axis. Visualization and measurement attempts have shown that the mixing layer growth in these jets leads to a fast destruction of the jet core, typically on a length of  the order of $\sim 4 L$. These jets also have  weaker gradients. Instead, in the overdense jets, the mixing layer spreads slowly, therefore the jet core has a longer path, usually $>4 L$. 

The global trend of the $\beta$ values can also be expressed by means of a general scaling law in the form
\bq
\beta = a + b\, \eta^n ,		
\label{scf}
\eq
which has the best fit on the experimental data with the coefficients $a=1.54,\,b=4.61,\,n=-0.32$, while the estimated error variance is 0.482.
The relevant curve $\beta(\eta)$, superimposed onto the experimental data, is shown in figure \ref{betafar}, 
and results to be very close
to model \rf{prfit}, except for very small values of $\eta$.

%

\section{Conclusions}

The present experiment has mainly focused on the effects of the jet-to-ambient density ratio on the morphology of strongly underexpanded jets,
on both short and long axial scales, in the range where compressibility is important. All the tested gases are monoatomic, and therefore there is no dependence of the stream properties on the specific heats ratio $\gamma$, which has simplified the characterization of the density effects. 
The Reynolds number in the nozzle, whihc is diameter based, ranges from $10^3$ to $5\cdot 10^4$, whereas the Reynolds number in the expansion, based on 
the axial length, ranges from $2 \cdot 10^3$ to $5\cdot 10^4$ for light jets and from $\cdot 10^5$ to $2\cdot 10^6$ for heavy jets in the same
Mach number interval, due to the different densities involved.
All the spreading angle measurements have been based on gas density and concentration values; the resulting mixing layer spreading dependence on the density ratios 
show that the jet/ambient mixing is in general more efficient in underdense jets. In particular, it has been shown that  the plumes in the near field
spread up to three times wider in underdense jets than in overdense ones; in the intermediate field  considered here, which is subsonic downstream from the Mach disk, the spreading of light jets is up to two times wider than for heavy jets. This different behaviour is slightly influenced by the Mach number in the near field, where it has been shown that increasing Mach numbers have a weak collimating effect, and reduce the plume width. The whole body of results has been synthesized by means of two simple scaling laws, which are valid in the near and in the intermediate fields, respectively.

\appendix
\section{\label{ALG}}

This section reports  details of the image post-processing algorithm used in this work, according to \cite{bel3}.
The method is based on equation \rf{intens}, i.e. the linear law $I=k\, n$ for the fluorescent emission $I$ of 
the excited gas where $n$ is the gas number density and $k$ a proportionality coefficient.

The emission $I$ is in fact a spectral superposition, therefore equation \rf{intens} also holds for the three colors 
(R, G, B) acquired by each pixel of the camera sensor and for any of their linear superposition:
\bq
R &=& k_R n \label{rint},\\
G &=& k_G n \label{gint},\\
B &=& k_B n \label{bint},\\
C &=&  (a k_R + b k_G + c k_B) n = k_C n \label{clin},
\eq
where $k_R, k_G, k_B$ are integral quantities that can be computed from known spectra or directly measured
(here they are obtained from experimental calibrations in pure gases at known temperatures). 

Relation \rf{clin} also holds for a two-gas mixture and, if the emission of the two gases is decoupled,  
the total intensity will be the sum of the individual ones. Labeling the ambient gas as $amb$ and the jet gas as 
$jet$,  emission $C$ can be written as the sum of two expressions of the kind \rf{clin}:
\bq
C = C_{amb} + C_{jet} = k_{amb} n_{amb} + k_{jet} n_{jet} 
\label{ctot}
\eq
and  again using equation \rf{clin} we obtain
\bq
k_C n = k_{amb} n_{amb} + k_{jet} n_{jet} .
\label{kc}
\eq
Dividing by the total numerical density $n$ and introducing the concentrations $z_{amb}=n_{amb}/n$ and $z_{jet}=n_{jet}/n$, equation \rf{kc} becomes
\bq
k_C = k_{amb} z_{amb} + k_{jet} z_{jet}, 
\label{kz}
\eq
where
\bq
k_{amb} &=& a k_{Ramb} + b k_{Gamb} + c k_{Bamb},\label{KA}\\
k_{jet} &=& a k_{R{jet}} + b k_{G{jet}} + c k_{Bjet}\label{KJ}.
\eq
Noting that the ratio between a pair of color intensities takes different values for different gases, it follows that 
two linear superpositions of the kind
\bq
   C_1=a_1 R + b_1 G + c_1 B  \label{C1}  \\
   C_2=a_2 R + b_2 G + c_2 B  \label{C2}
\eq
will usually give a ratio $C_1/C_2$ that varies with the species concentration. For a given pair of gases, a suitable choice of the six coefficients 
$a_i,b_i,c_i$ ($i$ = 1,2) will give a $C_1/C_2$ ratio that varies over the largest possible interval. This in turn leads to the concentration determination, if $C_1/C_2$ is rewritten using equations \rf{clin} and \rf{kz}:
\bq 
   \frac{C_1}{C_2}  = \frac{k_{C1}}{k_{C2}} = \frac{k_{amb_1} z_{amb} + k_{jet_1} z_{jet}} {k_{amb_2} z_{amb} + k_{jet_2} z_{jet} }.
\label{rel1}
\eq
The concentrations of the two gases are linked by $z_{amb}=1-z_{jet}$; writing $C_1/C_2 = r$,  
equation \rf{rel1} becomes 
\bq \label{rel2}
r = \frac{k_{amb_1} \left(1-z_{jet}\right) + k_{jet_1} z_{jet}} {k_{amb_2} \left(1-z_{jet}\right) + k_{jet_2} z_{jet} },
\eq
which can be solved with respect to $z_{jet}$, obtaining
\bq 
  z_{jet} = \frac{k_{amb_1} - k_{amb_2} r}{k_{amb_1} - k_{jet_1} + \left(k_{jet_2} - k_{amb_2}\right) r}. 
\eq
This gives  the concentration in the jet gas, for each pixel of the image, as a function of the ratio $r$ of the two superpositions $C_1, C_2$ read on the same pixel, provided that the four coefficients $k_{amb1},k_{amb2},k_{jet1},k_{jet2}$, defined through \rf{KA}-\rf{KJ}, are known through calibration.

Once the concentrations are known, the density can  easily be determined, as the total number density $n$ may be obtained from relations such as 
(\ref{clin}), e.g.  one of the followings relations can be used:
\bq
C_1 =  k_{C_1} n,  &\;\; &k_{C_1}=  k_{amb1} z_{amb} + k_{jet1} z_{jet}
\label{1clin} \\
C_2 =  k_{C_2} n, &\;\;  & k_{C_2}=  k_{amb2} z_{amb} + k_{jet2} z_{jet}.
\label{2clin}
\eq
Finally, the density is finally given by the equation
\bq
\rho = n (z_{amb} m_{amb} + z_{jet} m_{jet}) 
\eq
where $m_{amb}, m_{jet}$ are the molecular masses of the gases in the ambient and in the jet, respectively.


\end{document}